# Evaluation of reduced-graphene-oxide-supported gold nanoparticles as catalytic system for electroreduction of oxygen in alkaline electrolyte

Sylwia Zoladek[a], Iwona A. Rutkowska[a], Magdalena Blicharska[a], Krzysztof Miecznikowski[a], Weronika Ozimek[a], Justyna Orlowska[a], Enrico Negro[b], Vito Di Noto[b], Pawel J. Kulesza[a]*

[a] *Faculty of Chemistry, University of Warsaw, Pasteura 1, PL-02-093 Warsaw, Poland*
[b] *Department of Industrial Engineering, Università degli Studi di Padova in Department of Chemical Sciences, Via Marzolo 1, 35131 Padova (PD) Italy.*

*Corresponding author
E-mail: pkulesza@chem.uw.edu.pl; Tel: (+48) 22 5526344; Fax: (+48) 22 5526434

**Abstract**

Chemically-reduced graphene-oxide-supported gold nanoparticles with a diameter, of 40-60 nm are considered here as catalytic materials for the reduction of oxygen in alkaline medium in comparison to analogous systems based on conventional Vulcan carbon carriers. Gold nanoparticles are prepared by the chemical reduction method, in which the $NaBH_4$-prereduced Keggin-type phosphomolybdate heteropolyblue acts as the reducing agent for the precursor ($HAuCl_4$). Polyoxmetallate ($PMo_{12}O_{40}^{3-}$) capping ligands stabilize gold nanoparticle deposits, facilitate their dispersion and attachment to carbon supports. Indeed, it is apparent from the independent diagnostic voltammetric experiments (in 0.5 mol $dm^{-3}$ $H_2SO_4$) that heteropolymolybdates form readily stable adsorbates on nanostructures of both gold and carbon (reduced graphene oxide and Vulcan). It is reasonable to expect that the polyoxometallate-assisted nucleation of gold has occurred in the proximity of oxygenated defects existing on carbon substrates. Under conditions of electrochemical diagnostic experiments (performed in 0.1 mol $dm^{-3}$ KOH): (*i*) the phosphomolybdate adsorbates are removed from the interface as they undergo dissolution in alkaline medium; and (*ii*) the Au nanoparticles (Au loading, 30 μg $cm^{-2}$) remain well-dispersed on the carbon as evident from transmission electron microscopy. High electrocatalytic activity of the reduced-graphene oxide-supported Au nanoparticles toward reduction of oxygen in alkaline medium is demonstrated using cyclic and rotating ring-disk voltammetric experiments. Among important issues are possible activating interactions between gold and the support, as well as presence of structural defects existing on poorly organized graphitic structure of reduced graphene oxide (as evident from Raman spectroscopy).

# 1. Introduction

A tremendous amount of research has been carried out in the field of oxygen electroreduction, particularly with respect to potential applications in the fuel cell research [1-11]. Obviously, many efforts have been made to develop suitable alternative electrocatalysts efficient enough to replace electrocatalysts based on scarce strategic elements such as platinum-group metals (PGMs) [5,6,8,9,12,13]. Despite intensive research in the area of low-temperature fuel cells with proton-conducting electrolytes, the practical oxygen reduction catalysts still utilize systems based on PGMs.

More recently, the anion-exchange membrane low-temperature fuel cells [14,15], *i.e.* the systems mounting hydroxyl ($OH^-$)-conducting polymer membranes, have been proposed. Because they operate in alkaline environment, other elements beyond PGMs can act as reasonable candidates for implementation as the oxygen reduction catalytic sites [16]. Representative examples include gold, silver and nickel-based catalytic materials that are obviously much cheaper and more abundant, and are promising for future commercialization. In this respect, the systems based on gold nanoparticles capable of oxygen reduction in alkaline medium [17-20] should be mentioned. The enhanced catalytic activity of gold nanostructures is often attributed to existence of the high fraction of low-coordinated surface atoms at the corners and edges of Au nanoparticles, that play the roles of active sites [21,22]. The kinetics and mechanism of the oxygen electroreduction at Au(hkl) catalysts in alkaline medium is critically dependent on the interfacial gold structure as well as on the potential applied to the electrode [23]. It has been postulated that, in alkaline solutions, the four-electron reduction of oxygen on Au(100) involves the disproportionation mechanism [19] which seems to be enhanced on Au(100) in comparison to the performance of other two low-Miller index faces of Au. For example, on Au(111), the reduction of oxygen proceeds according to the 2-electron reaction mechanism with hydrogen peroxide as the final reduction product [24]. The role of the superoxide ($O_2^-$) adsorptive intermediate has been postulated to be more significant during the $O_2$-electroreduction in alkaline media relative to that in acidic media [19,23]. When it comes to application of carbon-supported gold nanoparticles (with a diameter of 7-8 nm) for oxygen reduction, their electrocatalytic activity seems to vary inversely with the size of the particles [17]. In this respect, the decreased hydrophilicity and coordination of the smaller clusters of gold is believed to increase to the energy of binding of between O and Au and to decrease the activation energy during the dioxygen dissociative chemisorption.

The $O_2$-reduction electrocatalysts are typically nanocomposite materials utilizing metal nanoparticles bearing the active sites dispersed on suitable supports. While exhibiting long-term stability, a useful support should facilitate dispersion, provide easy access of reactants, and assure good electrical contact with active sites. In spite of limitations related to the durability, carbon nanoparticles of approximately 20-50 nm diameters (e.g. Vulcan XC-72R) are commonly utilized as supporting materials. Because of the high specific surface area and excellent thermal, mechanical and electrical properties, graphene and graphene-based materials [25-27] have recently been considered as supports for catalysts [28-32]. Under such conditions, the parasite effects related to agglomeration and thus degradation of catalytic nanoparticles are likely to be largely prevented.

In the present work, we consider the chemically reduced graphene-oxide-supported gold nanoparticles as catalytic systems for the electroreduction of oxygen in basic medium. The electrocatalytic diagnostic experiments involve comparative measurements utilizing commonly-used Vulcan (carbon) supports as carriers for the analogous gold nanoparticles. Here application of inorganic Keggin-type heteropolymolybdates ($PMo_{12}O_{40}^{3-}$) as capping ligands (capable of chemisorbing on both gold and carbon substrates [33-40] facilitates deposition, nucleation, stabilization and thus controlled growth of gold nanoparticles on surfaces of both Vulcan and graphene nanostructures. In the latter case, we have utilized the so-called reduced graphene oxide which, contrary to conventional graphene, still contains oxygen functional groups regardless of subjecting it to the chemical reduction step [25-27]. By analogy to graphene oxide, the existence of oxygen groups in the plane of carbon atoms of reduced graphene oxide not only tends to increase the interlayer distance but also makes the layers somewhat hydrophilic. Furthermore, during fabrication of the catalytic systems (in acid medium), the adsorbed polymolybdates [9,34] are likely to bind gold nanoparticles via the oxygen or hydroxyl groups on graphene and Vulcan surfaces. During operation in alkaline medium, polyoxometallates decompose and disappear; but the catalytically active gold centers remain stable on carbon surfaces. It is apparent from the diagnostic cyclic voltammetric and rotating ring-disk measurements in the oxygen-saturated 0.1 mol dm$^{-3}$ KOH that the system utilizing Au catalyst deposited on the chemically reduced graphene oxide has exhibited higher electrocatalytic currents (and produced lower amounts of the undesirable hydrogen peroxide intermediate) during oxygen reduction (relative to the performance of gold nanoparticles deposited on conventional Vulcan-carbon support or simple bare Au nanoparicles). The enhancement effect is particularly sound in the high potential range (0.8-1.0 V vs. RHE). On the whole, the combined effect of the high surface area and electrical conductivity of reduced graphene oxide should contribute to the overall enhancement effect and activity of Au nanoparticles toward the oxygen reduction.

## 2. Experimental

Chemicals were commercial materials of the highest available grade, and they were used as received. KOH, HCl, HNO$_3$, K$_2$SO$_4$, ethanol and K$_3$[Fe(CN)$_6$] were obtained from POCh (Poland). The hydrogen tetrachloroaurate (III) trihydrate, HAuCl$_4$·3H$_2$O (>99.9%); sodium borohydride (powder, 98%), NaBH$_4$; phosphomolybdic acid hydrate, H$_3$PMo$_{12}$O$_{40}$·nH$_2$O (ACS. reagent), 5 wt% Nafion solution were purchased from Sigma-Aldrich and were used without any further purification. Nitrogen and oxygen gases (purity 99.999%) were from Air Products (Poland).

Graphene oxide sheets of 300-700 nm sizes (thickness, 1.1±0.2 nm) were from Megantech. Reduced graphene oxide (rGO) was obtained using sodium borohydride as reducing agent at 80˚C according to the procedure described earlier [41].

Vulcan XC-72R conductive carbon black was purchased from Cabot and purified by treatment in 12 mol dm$^{-3}$ HCl solution for 1h followed by washing out with distilled-deionized water several (3-4) times to remove any possible leftovers. The resulting carbon black particles were functionalized with interfacial oxygen-containing groups by subjecting them to 3 mol dm$^{-3}$ HNO$_3$ for 8 h under reflux conditions, then washed thoroughly with water, (collected under high-speed centrifuging conditions) and, finally, dried completely at 60°C. Solutions (including 0.1 mol dm$^{-3}$ KOH) were prepared by dissolving the respective chemical in the triply-distilled subsequently-deionized (with Milli-Q water purification system) water.

Fabrication of gold nanoparticles modified with heteropolymolybdates (PMo$_{12}$O$_{40}^{3-}$) was achieved in a manner analogous to the approach described in our previous papers [34-38]. First, a stoichiometric volume of the freshly prepared aqueous 0.016 mol dm$^{-3}$ sodium tetrahydroborate (NaBH$_4$) was added to the 30 cm$^3$ of o 2.56 mmol dm$^{-3}$ aqueous solution of phosphododecamolybdic heteropolyacid (H$_3$PMo$_{12}$O$_{40}$) under vigorous stirring. In the next step, an equivalent volume (3 ml) of the aqueous 7.5 mmol dm$^{-3}$ HAuCl$_4$ (precursor) was added. The resulting colloidal suspension was centrifuged, the supernatant solution was removed and, subsequently, replaced with water. A final solution of the water-dispersed gold nanoparticles (stabilized by PMo$_{12}$O$_{40}^{3-}$) had a volume of 0.5 cm$^3$.

The syntheses of phosphomolybdate-modified gold nanoparticles supported onto Vulcan XC72R carbon and reduced graphene oxide matrices were performed in the analogous manner but in the presence of an appropriate carbon support. To assure good dispersion of nanostructured carbon supports in water, they have been pre-modified with PMo$_{12}$O$_{40}^{3-}$ adsorbates, namely, by exposing either 0.01 g of purified Vulcan XC72R carbon or chemically reduced graphene-oxide to 30 cm$^3$ of the aqueous 2.56 mmol dm$^{-3}$ phosphododecamolybdic acid (H$_3$PMo$_{12}$O$_{40}$) solution. The latter functionalization procedures were facilitated by prolonged mixing (12 h). Then, the stoichiometric volume of freshly prepared aqueous 0.016 mol dm$^{-3}$ sodium tetrahydroborate (NaBH$_4$) was added to the phosphomolybdate-functionalized carbon supports in order to transform the oxidized $H_3PMo_{12}O_{40}$ adsorbates into partially reduced $H_3[H_4P(Mo^V)_4(Mo^{VI})_8O_{40}]$ heteropolyblue forms. To obtain a gold loading on the level of 30 wt% of Au on the appropriate heteropolyblue-modifed carbon, an equivalent volume of the aqueous 7.5 mmol dm$^{-3}$ chloroauric acid ($HAuCl_4$) solution was added to the respective suspension. The solutions were mixed for 8 h at room temperature to ensure the reaction completeness. The phosphomolybdate adsorbates are expected to exhibit attractive interactions between carbon supports and growing gold nuclei during the deposition steps. Excessive amounts of phosphomolybdate anions were removed by consecutive centrifuging steps involving exchanges of supernatant solutions with water (repeated at least three times). The final solutions of gold nanoparticles supported onto chemically-reduced graphene-oxide (Au/Graphene) or onto Vulcan carbon (Au/Vulcan) were adjusted to the volume of 0.5 cm$^3$.

All electrochemical experiments were performed in a conventional three-electrode cell using CH Instruments (Austin, TX, USA) Models: 600B and 750A workstations. A Saturated calomel electrode (SCE) was the reference electrode; it exhibited potential of *ca.* 240 mV relative to the reversible hydrogen electrode (RHE). The counter electrode was a carbon wire. A rotating ring disk electrode (RRDE) working assembly was from Pine Instruments; it included a glassy carbon (GC) disk and a Pt ring. The radius of the GC disk electrode was 2.5 mm; and the inner and outer radii of the ring electrode were 3.25 and 3.75 mm, respectively. As a rule, potentials presented in this study were recalculated and expressed against the reversible hydrogen electrode (RHE). The collection efficiency (N) equal to 0.39 was determined from the ratios of ring and disk currents obtained from six independent RRDE measurements performed at the rotation rate of 1600 rpm using the argon-saturated 0.005 mol dm$^{-3}$ $K_3[Fe(CN)_6]$ in 0.01 mol dm$^{-3}$ $K_2SO_4$ solution [42].

Prior to modification the glassy carbon electrodes were activated by polishing on a cloth wetted with successively finer grade aqueous alumina slurries (grain sizes from 5 to 0.5 μm). All measurements were carried out at room temperature 22±2°C.

Morphology of samples was assessed using Libra Transmission Electron Microscopy120 EFTEM (Carl Zeiss) operating at 120 kV.

Visible absorption spectra (400-800 nm) of the following aqueous suspensions were recorded using UV-Vis Lambda Spectrophotometer (Perkin Elmer): dispersions of chemically-reduced graphene-oxide modified with heteropolymolybdate ($H_3PMo_{12}O_{40}$) or partially reduced heteropolyblue ($H_3[H_4P(Mo^V)_4(Mo^{VI})_8O_{40}]$), and of the phosphomolybdate-functionalized chemically-reduced graphene-oxide onto which gold nanoparticles were deposited.

The Raman spectra were collected with a confocal Raman Microscope (model DRX, Thermo Scientific) and using an excitation laser with a wavelength of 532 nm.

The infrared spectra, regular in KBr of $H_3PW_{12}O_{40}$ and by reflectance (IRRAS) of $H_3PW_{12}O_{40}$ adsorbate (deposited onto gold-covered glass) were taken with Shimadzu 8400 FTIR spectrometer (where needed using Specular Reflectance Accessory Model 500 provided by Spectra Tech). An $80^o$-beam incidence angle (with respect to the surface normal) was utilized.

A catalytic film of gold nanoparticles was deposited on the glassy carbon disk electrode according to the following procedure: a 220 µl aliquot from the final solution (0.5 cm$^3$) of the water-dispersed phosphomolybdate-stabilized gold was suspended in 240 µl of ethanol and then subjected to vigorous stirring for 1 h. Later, 40 µl of 5% Nafion water-alcoholic solution (from Aldrich) was added; the resulting mixture was left under magnetic stirring for 6 h. Finally, 2 µl of the resulting suspension of gold nanoparticles was dropped onto the surface of glassy carbon disk (RRDE); the suspension was air-dried at room temperature.

A catalytic ink containing gold nanoparticles supported onto chemically- reduced graphene-oxide (Au/rGO) or Vulcan carbon (Au/Vulcan) were prepared according to the procedure as follows. The 220 µl aliquots from the appropriate final dispersions of Au/rGO or Au/Vulcan were introduced to 240 µl ethanol samples and, subsequently, subjected to 24 h stirring to obtain homogenous mixture. Then 40 µl of Nafion (5% alcoholic solution) was added to the suspension. The resulting solution had been magnetically stirred for 24 h; and, later, subjected to drying in air at room temperature for 60 min. As a rule, appropriate amounts of the resulting catalytic inks were dropped onto surfaces of glassy carbon electrodes to obtain loadings of gold nanoparticles equal to 30 µg cm$^{-2}$. For comparison, inks containing only bare (purified) Vulcan XC-72R carbon black or chemically- reduced graphene-oxide were prepared according to analogous procedures, except in the absence of gold nanoparticles. In other words, 4.6 mg of Vulcan carbon or chemically- reduced graphene-oxide was suspended in the 460 µl of ethanol and 40 µl of Nafion (5% alcoholic solution). Later 2 µl of the appropriate bare carbon suspension was introduced onto the surface of the glassy carbon disk electrode; and the suspension was air-dried at room temperature.

The catalytic films were activated by performing 50 full voltammetric potential cycles in the range from -0.1 to 1.3 V vs. RHE at 50 mV s$^{-1}$ until stable voltammetric responses were observed. Following medium transfer to alkaline solution, the phoshomolybdate capping layers were removed. Typical cyclic voltammograms were recorded by scanning potential from 0.1 to 1.4 V in 0.1 mol dm$^{-3}$ KOH at a scan rate of 50 mV s$^{-1}$. Nevertheless, the first voltammogrms were recorded at 10 mV s$^{-1}$ under nitrogen atmosphere to obtain the representative background responses. Later, the oxygen reduction measurements were done in the oxygen-saturated 0.1 mol dm$^{-3}$ KOH. All RRDE polarization curves were recorded at the scan rate of 10 mV s$^{-1}$, typically with a rotation rate of 1600 rpm. The ring potential was maintained at 1.21 V *vs.* RHE to oxidize any hydrogen peroxide produced. Prior to each experiment, the electrolyte solution was bubbled for 30 min with $O_2$ and $N_2$, respectively. A constant nitrogen (or oxygen) flow over the solution was maintained during all measurements.

## 3. Results and Discussion

*3.1. Physicochemical Identity of Reduced Graphene Oxide*

Graphene oxide, GO, and partially reduced graphene oxide, rGO contain various carbon–oxygen groups (hydroxyl, epoxy, carbonyl, carboxyl), in addition to the large population of water molecules still remaining in the reduced samples. Independent elemental analysis based on the C 1s and O 1s XPS spectra from XPS measurements [43] showed that the oxygen content in rGO was in the range from 8.6 to 12.1 at%; the C-to-O ratio was on the level of 7.1–10.3. When compared to the analogous parameters of the commercially available GO, the oxygen content and the C-to-O ratio values were more than three times lower and more than three times higher, respectively [43].

The rGO material has also been characterized (relative to GO) by Raman spectroscopy (Fig. 1). The assignment of the Raman bands has been carried out in accordance with the previous reports [44,45]. As expected, the Raman spectra of Fig. 1 show two large peaks in the range of 1300-1600 cm$^{-1}$: one peak near 1350 cm$^{-1}$, which stands for the D band originating from the amorphous structures of carbon, and the second one close 1580 cm$^{-1}$, which is correlated with the G band and reflects the graphitic structures of carbon. It comes upon comparison of curves a (GO) and b (rGO) in Fig. 1 that the intensities of G and D bands in rGO, relative to the analogous bands in GO, are lower and higher, respectively. This result implies presence of interfacial defects as well as the lower degree of organization of the graphitic structure of rGO relative to GO. These features could lead to the enhancement effects in electrocatalysis.

It has been established that polyoxometallates, in particular the Keggin-type phosphododecamolybdic acid, $H_3PMo_{12}O_{40}$ ($PMo_{12}$), adsorb readily on many solid surfaces including noble metal nanoparticles and carbon nanostructures. Due to existence of the surface defects and the oxygen-containing functional groups on rGO, $PMo_{12}$ is expected to undergo adsorption and to form electroactive adsorbates on rGO as well. Fig. 2A shows a cyclic voltammogram of $PMo_{12}$-modified rGO which was spontaneously deposited on bare glassy carbon through dipping of the electrode substrate in the aqueous (unbound $PMo_{12}$-free) colloidal suspension of PMo$_{12}$-modified rGO for 20 min. In view of our previous reports [34-38], the observed redox transitions (pairs of surface voltammetric peaks in Fig. 2A) reflect three consecutive two-electron processes:

$$PMo^{VI}_{12}O^{3-}_{40} + ne^- + nH^+ \Leftrightarrow H_nPMo^V_nMo^{VI}_{12-n}O^{3-}_{40} \qquad (1)$$

where *n* is equal to 2, 4 or 6. It is noteworthy that the full-width values at half-maximum of the second oxidation or reduction peaks (appearing at about 0.2 V in Fig. 2A) are very close to the theoretical 45 mV (expected for an ideal two-electron surface-type voltammetric peak). The peak heights of Fig. 2A have been found to be directly proportional to scan rates up to at least 500 mV s$^{-1}$. Furthermore, under such conditions, the formal potentials of three sets of the PMo$_{12}$ surface-peaks (Fig. 2A) have been practically independent of the scan rate.

Presence of the adsorbed phosphododecamolybdates (PMo$_{12}$) on rGO was also confirmed by performing *ex-situ* FTIR measurement on the PMo$_{12}$-modified rGO platelets deposited onto the conventional gold-coated glass slide substrate (Fig. 2B. Curve b). A conventional spectrum of PMo$_{12}$ in KBr (Fig. 2B, Curve a) is also provided. As before in the cases of PMo$_{12}$-modifed gold, platinum and carbon nanostructure [9,33-40], clear analogy between the infrared spectrum of the PMo$_{12}$ adsorbate (on rGO) and the model one should be mentioned. The fact that positions of the P-O-Mo stretching modes (1064 cm$^{-1}$) are virtually identical (Fig. 2B), indicates that adsorption of PMo$_{12}$ on rGO does not affect the inner P-O bonding in the polyoxometallate structure. Certain shifts (increases) of frequencies of the Mo-O-Mo stretching modes (related to "corner" and "edge" oxygens) from 787 and 869 to 804 and 872 cm$^{-1}$, respectively, reflect most likely strengthening and shortening of the Keggin-type PMo$_{12}$O$_{40}^{3-}$ structures upon adsorption (immobilization) on the rGO surface. A shift of the external (terminal) Mo-O bond frequency from 961 (Fig. 4, Curve a) to 964 cm$^{-1}$ and related structural changes are very minor in this respect. As before for PMo$_{12}$ adsorbates at different surfaces [9,33-40], the present data are consistent with the view that attachment or interfacial interactions of PMo$_{12}$ with rGO (defects and/or oxygen functional groups) involve mostly "corner" oxygen atoms. Remembering that $PMo_{12}$ adsorbs strongly on gold nanoparticles [35,36], it is reasonable to expect that phosphomolybdates facilitate deposition and distribution of Au nanoparticles on carbon supports including rGO.

*3.2. Formation and morphology of dispersed gold nanoparticles*

The $PMo_{12}$-modified Au nanoparticles (supported and unsupported) were characterized using Transmission Electron Microscopy (TEM). It is apparent from Fig. 3 that while unsupported gold particles have diameters that are fairly uniform in the range between *ca.* 30 and 40 nm diameters, the carbon-supported (both onto rGO and Vulcan) particles - though slightly larger and less uniform - have comparable sizes typically ranging from 30 to 50 nm. On mechanistic grounds, gold nucleation may occur at carbons' "defect" sites, including surface polar groups and polyoxometallate adsorbates. It is reasonable to expect that the partially reduced (heteropolyblue) $PMo_{12}O_{40}^{3-}$ sites induce generation of somewhat larger gold nanoparticles (relative to unsupported Au; compare Figs 3b and 3c to Fig. 3a).

We have utilized conventional spectrophotometry in visible region to monitor formation of phosphomolybdate-modified gold nanoparticles immobilized onto chemically reduced graphene (Fig. 4). Addition of sodium tetrahydroborate to the solution of $H_3PMo_{12}O_{40}$ and rGO results in a rapid change of color from brown (Fig. 4a, dotted line) to dark blue (Fig. 4b, dashed line) thus indicating formation of the heteropolyblue form of phosphododecamolybdate. A low-energy broad tail extending from 500 to well-beyond 800 nm (Fig. 4b) is consistent with the existence of the inter-valence charge-transfer between Mo(V) and Mo(VI) centers [46-50]. Then addition of the $HAuCl_4$ precursor to the partially-reduced mixed-valence Mo(VI,V) heteropolyblue solution has largely depleted the portion of spectrum assigned as the inter-valence charge-transfer (Fig. 4c, solid line). The result is consistent with the re-oxidation of Mo(V) to Mo(VI) that accompanies reduction of Au(III) in $HAuCl_4$ to metallic Au. The appearance of new absorption band at about 540 nm (Fig. 4c) should be correlated with the typical plasmonic resonance band characteristic of spherical gold nanoparticles [50-56].

*3.3 Electrochemical identity of Au nanoparticles*

To comment on the influence of chemically-reduced graphene-oxide (rGO), relative to Vulcan carbon, nanostructures on the electrochemical characteristics of gold nanoparticles, the voltammetric experiments (Fig. 5) have been performed on the following gold nanoparticles: (A) unsupported Au (as for Fig. 5A), (B) rGO-supported Au (as for Fig. 5B), and (C) Vulcan-supported Au (as for Fig. 5C). The dotted lines in Figs 5B and 5C stand for the responses of Au-free rGO and Vulcan carbons. Relative to the performance of bare Au nanoparticles (Fig. 5A), introduction of rGO and Vulcan supports (Figs 5B and 5C) leads to increase of background currents originating from the double-layer-type charging/discharging effects occurring on the large-surface-area carbon surfaces. Under such conditions, the responses characteristic of redox transitions leading to the formation and reduction of gold oxides, that appear in potential range from 1 V to 1.4 V (Fig. 5A), would be obviously somewhat hidden within relatively high background currents of carbon supports mentioned above. The fact that these background currents are even higher in the presence of dispersed Au nanoparticles could be explained in terms of improved charge distribution following incorporation of Au nanostructures. Strong interactions between gold nanoparticles and chemically-reduced graphene-sheets involving electron transfers from underlying graphene to Au sites were postulated before [56]. Our present voltammetric results of Fig. 5 imply some changes in the systems' characteristics, including certain current enlargement effects, during re-reduction of Au oxides on rGO, relative to Vulcan or bare Au nanoparticles. Future research, that would include systematic surface (XPS and Raman) measurements, is needed along this line.

*3.4 Electroreduction of oxygen*

It is apparent from the diagnostic linear scan voltammetric experiments that oxygen reduction proceeds at gold nanoparticles at fairly positive peak potentials, namely at ca. 0.8 V. (Fig. 6). But the electrooxidation currents (net or background subtracted) have been found to be dependent on the application and choice of carbon support though sizes and distribution of Au nanoparticles are, in all cases (Fig. 3), comparable. For comparison, the activities of carbon (rGO, Vulcan) supports themselves toward reduction of oxygen under analogous voltammetric conditions are illustrated in a form of dotted lines (in Figs 6B and 6C).

The voltammetric data of Fig. 6 is consistent with the relatively highest catalytic activity of the rGO-supported Au-nanoparticles during electroreduction of oxygen in 0.1 mol dm$^{-3}$ KOH. Here it should be noted that the rGO-support itself exhibits appreciable activity toward $O_2$-reduction but at more negative potentials (Fig. 6B; dotted line). Indeed the respective voltammetric peak is shifted more than 250 mV toward more negative values. In other words, at potentials where the main voltammetric peak (characteristic of rGO-supported Au) is developed (at ca. 0.8 V in Fig. 6B), the contribution from the rGO-support is negligible (compare dotted and solid lines in Fig. 6). In this respect, rGO seems to act as highly conductive robust support exhibiting both hydrophobic and hydrophilic domains that facilitate stability and mobility of charge (e.g. $OH^-$ ions), respectively, at the electrocatalytic interface. Roles of possible metal (Au) – support (rGO) interactions, as postulated earlier [56], in addition to interfacial structural defects on rGO, as apparent from Raman measurements (Fig. 1), would require additional future investigations.

Similar observations concern electroreduction of oxygen at Vulcan-supported Au-nanoparticles (Fig. 6C). As in a case of rGO-supported Au (Fig. 6B), application of highly conductive carbon currier has led to the enhancement of the electrocatalytic performance of the Au nanoparticles, as demonstrated by the positive shift of the peak potential from 0.77 to 0.8 V accompanied by an enlargement of the peak current (Fig. 6C), when compared to the performance of unsupported Au bare gold nanoparticles (Fig. 6A). Explanation of these phenomena should take into account possibility of better distribution of gold active sites upon deposition onto carbon support and improved charge distribution at the interface. Nevertheless, the application of rGO has resulted in the highest catalytic activity toward electroreduction of oxygen (Fig. 6B). When compared to the Fig. 6C data (Vulcan support), more than 20% higher voltammetric current density (net $O_2$-reduction) has been observed upon application of rGO. It can be concluded that, in comparison to the commonly utilized Vulcan carriers, rGO seems to be a promising nanostructured material for supporting catalytic metal nanoparticles.

*3.4. Diagnostic RRDE experiments*

To get insight into the mechanism and dynamics of the electroreduction of oxygen at rGO-supported Au-nanoparticles, relative to the system's behavior at unsupported and Vulcan-supported Au, the rotating ring-disk electrode (RRDE) voltammetric experiments have been performed.

Figure 7 illustrates representative disk (voltammetric) and simultaneous ring (upon application of 1.21 V) steady-state currents recorded during the reduction of oxygen (in the $O_2$-saturated 0.1 mol dm$^{-3}$ KOH at 1600 rpm rotation rate and 10 mV s$^{-1}$ scan rate) using (A) unsupported, (B) rGO-supported, and (C) Vulcan-supported Au nanoparticles. Upper curves in Figs 7A-7C stand for current responses at Pt ring, and they correspond to the relevant disk electrode current responses. The ring current values shall be attributed to the oxidation of $HO_2^-$ intermediate generated during the reduction of oxygen at the disk electrode. Under hydrodynamic voltammetric conditions of Fig. 7, while the disk current densities are roughly comparable for all Au-containing systems, the different ring currents have been produced thus implying formation of distinct amounts of the undesirable $HO_2^-$ intermediate. Also judging from shapes of disk and ring current-potential curves, it can be rationalized that the systems operate according to mixed two/four electron mechanisms and produce different amounts of $HO_2^-$ at different potentials. The fact that no plateau currents are formed in both the disk and ring responses implies further reduction of the $HO_2^-$ intermediate to water at lower potentials.

It is noteworthy that ring currents corresponding to the oxidation of $HO_2^-$ are much lower for rGO-supported Au nanoparticles (Fig. 7B) relative to the analogous responses for the unsupported Au and Vulcan-supported Au (Figs 7A and 7C). The results described here (Fig. 7B) are almost perfectly reproducible (currents within 5%) during prolonged experimentation (for at least 10h). Furthermore, simple examination of the dotted lines in Figs 7B and 7C (standing for RRDE responses of Au-free rGO and Vulcan supports, respectively) shows that, contrary to the rGO-supported Au (where Au nanoparticles, rather than rGO, seem to exhibit electrocatalytic activity during $O_2$-reduction), Vulcan itself is electrocatalytic toward $O_2$-reduction in alkaline medium. In other words, rGO seems to act as robust largely inert carrier for catalytic Au nanoparticles. Finally, careful examination of the systems' disk currents at potentials higher than 0.8 V (Fig. 8) clearly implies that the rGO-supported Au nanoparticles start to drive electroreduction of oxygen at potentials as high as 0.9 V (which is neither a case for unsupported nor Vulcan-supported Au nanoparticles). This result, when correlated with the relatively unique cyclic voltammetric response of the rGO-supported Au recorded in the $O_2$-free KOH electrolyte (Fig. 5) may imply existence of specific activating interactions between Au and rGO [56]. Further research is needed in this respect.

Figure 9 illustrates the percent amount of $H_2O_2$ (%$H_2O_2$) formed during reduction of oxygen under the conditions of RRDE voltammetric experiments of Fig. 9. The actual calculations have been done using the equation given below [42,58]:

$$\%_{H_2O_2} = \frac{200 I_{ring}/N}{I_{disk} + I_{ring}/N} \qquad \text{(Equation 1)}$$

where $I_{ring}$ and $I_{disk}$ are the ring and disk currents, respectively, and N is the collection efficiency. The results clearly show that the production of $H_2O_2$ is the lowest for system utilizing gold nanoparticles supported onto chemically-reduced graphene-oxide (Fig. 9A).

The overall number of electrons exchanged per $O_2$ molecule (n) was calculated as a function of the potential using the RRDE voltammetric data of Fig. 10 and using equation [58,59]:

$$n = \frac{4I_{disk}}{I_{disk} + I_{ring}/N} \quad \text{(Equation 2)}$$

The corresponding number of transferred electron (n) per oxygen molecule involved in the oxygen reduction was estimated to be as follows: (A) 2.94-3.20, (B) 3.37-3.60, and (C) 3.03-3.25 for the oxygen reduction at unsupported, rGO-supported, and Vulcan supported Au nanoparticles. The results demonstrate that, while the approximately three-electron reduction mechanism (with some formation of $HO_2^-$ intermediate) is the dominating pathway for the oxygen reduction (in alkaline medium) at all three types of Au-containing catalytic systems, utilization of the chemically-reduced graphene-oxide based supports produces the hybrid system exhibiting the relatively highest activity toward electroreduction of oxygen in alkaline medium.

## 4. Conclusions

This study clearly demonstrates that the chemically-reduced graphene-oxide acts as a robust and activating support for dispersed gold nanoparticles during electrocatalytic reduction of oxygen in alkaline medium (0.1 mol dm$^{-3}$ KOH). Among important issues is application of Keggin-type phosphododecamolybdates capable of strongly adsorbing on metal and carbon surfaces and acting as both capping (stabilizing Au nanostructures) and linking (to the largely defected reduced graphene oxide supports) agents during preparation of gold nanoparticles in acid medium. To achieve this goal, $H_3PMo_{12}O_{40}$ has been pre-reduced with $NaBH_4$ to heteropolyblue thus driving reductive deposition (from $HAuCl_4$) of well-dispersed Au nanoparticles. Later, during experiments in alkaline medium, polyoxometallates adsorbates are removed readily from the electrocatalytic interface.

For the same loadings of catalytic gold nanoparticles (30 μg cm$^{-2}$), application of the reduced graphene oxide support results in formation of lower amounts of the undesirable HO$_2^-$ intermediate (during O$_2$-reduction in alkaline medium) relative to the performance of unsupported or Vulcan-carbon-black supported Au catalysts under analogous conditions. Moreover the onset potential for the oxygen reduction has been the most positive (0.9 V) in a case of the system utilizing reduced graphene oxide. It should be mentioned here that the synergistic effects and activating interactions between catalytic metal nanoparticles and nanostructured graphene supports were postulated before [56,57,59] with respect to lowering the dissociation activation energy for molecular O$_2$ through accelerating the charge transfer from metal in presence of graphene and by reducing stability of the HO$_2^-$ intermediate species. Because the number of transferred electron per oxygen molecule was still below four (needed for the reduction of O$_2$ directly to H$_2$O), our future work will not only concentrate on elucidating and optimizing of the activating phenomena but on intentional decorating of the system with trace amounts of highly catalytic noble metal nanoparticles [60].

**Acknowledgments**

This work was supported by the European Commission through the Graphene Flagship – Core 1 Project (GA-696656). The Polish side appreciates support from National Science Center (Poland) under Maestro Project 2012/04/A/ST4/00287 and from Ministry of Science and Higher Education, Poland. Financial support for Sylwia Zoladek from Foundation for Polish Science under Start Project is also acknowledged.

**Figure captions**

**Fig. 1.** Raman spectra of **(a)** graphene oxide (GO) and **(b)** chemically-reduced graphene oxide (rGO).

**Fig. 2. (A)** Cyclic voltammetric response of $H_3PMo_{12}O_{40}$ adsorbate on reduced graphene oxide (electrolyte: 0.5 mol dm$^{-3}$ $H_2SO_4$; scan rate: 10 mV s$^{-1}$); **(B)** FTIR spectra of **(a)** $H_3PMo_{12}O_{40}$ in KBr and **(b)** by reflectance of $H_3PMo_{12}O_{40}$ adsorbate on reduced graphene oxide (deposited on gold-covered glass substrate).

**Fig. 3.** Transmission electron microscopic images of **(A)** unsupported gold nanoparticles, **(B)** reduced-graphene-oxide (rGO) supported Au nanoparticles, and **(C)** Vulcan-carbon-black supported Au nanoparticles. Histograms **A'**, **B'**, and **C'** display distribution of sizes of the respective gold nanostructures.

**Fig. 4.** Visible absorbance spectra of the following solutions (suspensions): **(a)** $H_3PMo_{12}O_{40}$ with chemically-reduced graphene-oxide (rGO) nanostructures; **(b)** partially reduced $H_3PMo_{12}O_{40}$ (heteropolyblue) and rGO nanostructures, and **(c)** gold nanoparticles supported onto $H_3PMo_{12}O_{40}$-modified RGOreduced-graphene-oxide nanostructures.

**Fig. 5.** Cyclic voltammetric responses of **(A)** unsupported gold nanoparticles, **(B)** reduced graphene oxide (rGO) supported gold nanoparticles, and **(C)** Vulcan-carbon-black supported gold nanoparticles. Electrolyte: 0.1 mol dm$^{-3}$ KOH. Scan rate: 50 mV s$^{-1}$. Gold loading: 30 μg cm$^{-2}$.

**Fig. 6.** Background-subtracted linear scan voltammetric responses recorded for the reduction of oxygen at the following catalytic layers: **(A)** unsupported gold nanoparticles, **(B)** reduced

graphene oxide (rGO) supported gold nanoparticles, and **(C)** Vulcan-carbon-black supported gold nanoparticles. Dotted lines stand for responses of the respective Au-free carbon supports. Electrolyte: oxygen-saturated 0.1 mol dm$^{-3}$ KOH. Scan rate: 10 mV. Gold loading: 30 μg cm$^{-2}$.

**Fig. 7.** Normalized (background subtracted) rotating ring-disk voltammograms for oxygen reduction at **(A)** unsupported gold nanoparticles, **(B)** reduced graphene oxide (rGO) supported gold nanoparticles, and **(C)** Vulcan-carbon-black supported gold nanoparticles. Dotted lines stand for responses of the respective Au-free carbon supports. Electrolyte: oxygen-saturated 0.1 mol dm$^{-3}$ KOH. Scan rate: 10 mV s$^{-1}$. Rotation rate: 1600 rpm. Ring currents are recorded upon application of 1.21 V. Gold loading: 30 μg cm$^{-2}$.

**Fig. 8.** Normalized (background subtracted) rotating disk voltammograms for the oxygen reduction recorded recorded in the range of high potentials (from 1.0 to 0.8 V) using the electrocatalytic systems described as for Fig. 7.

**Fig. 9.** Percent fraction of hydrogen peroxide (% $H_2O_2$) produced during electroreduction of oxygen (and detected at ring at 1.21 V) under conditions of the RRDE voltammetric experiments as for Fig. 7.

**Fig. 10.** Numbers of transferred electrons (n) per oxygen molecule during electroreduction of oxygen under conditions of the RRDE voltammetric experiments as for Fig. 7.

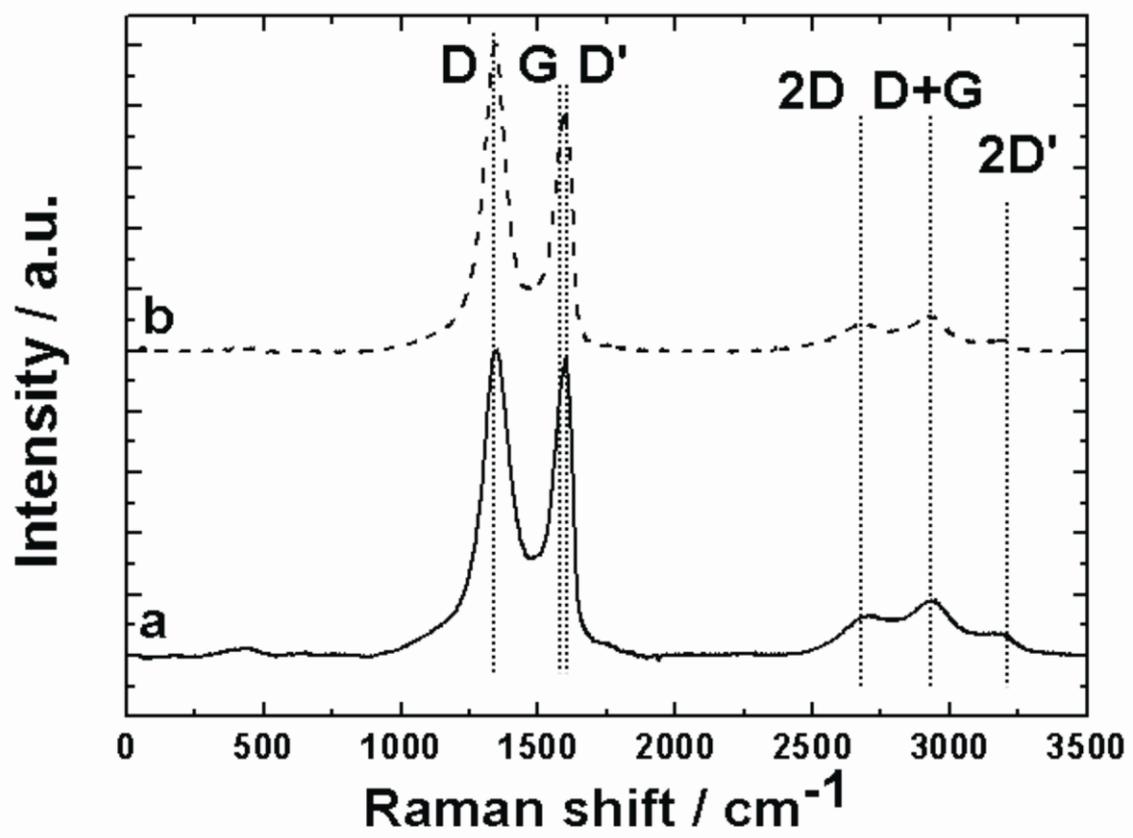

Fig. 1

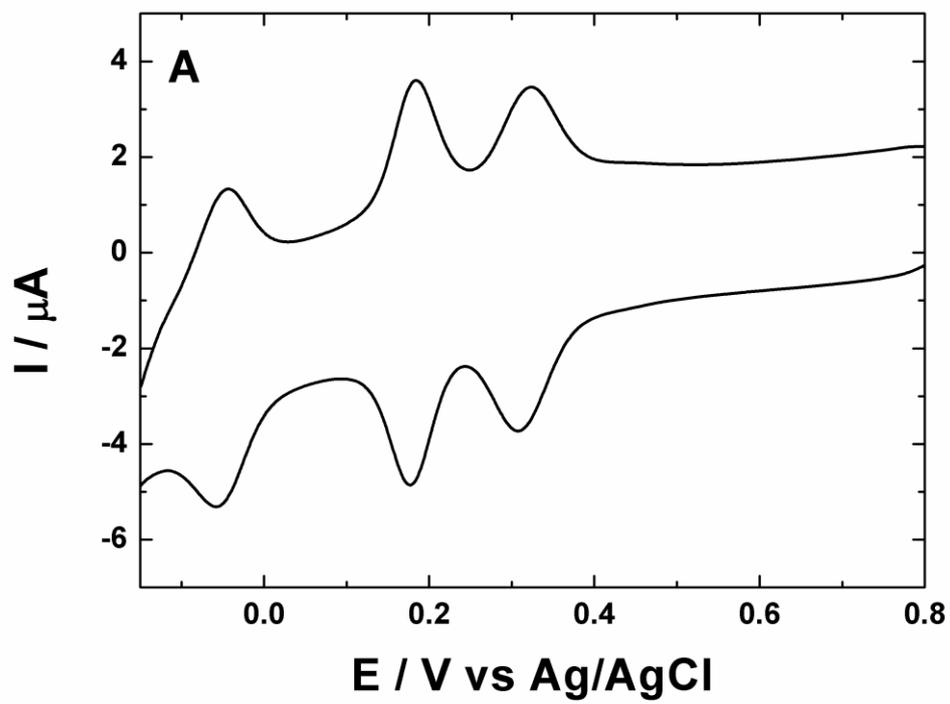

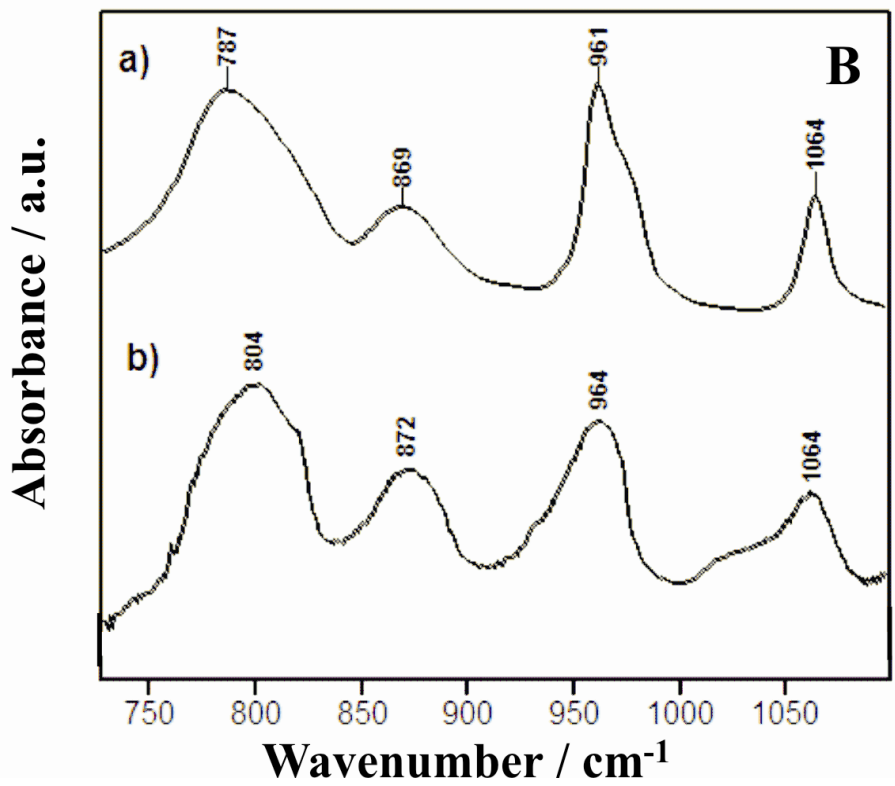

Fig 2

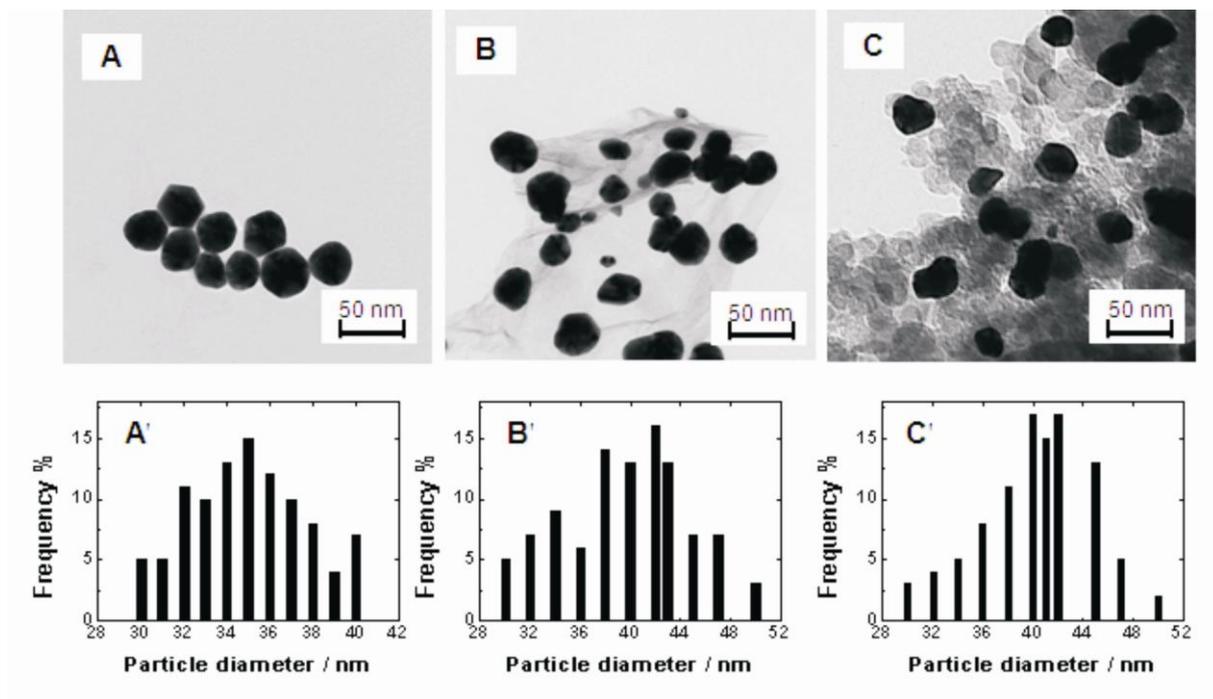

Fig. 3

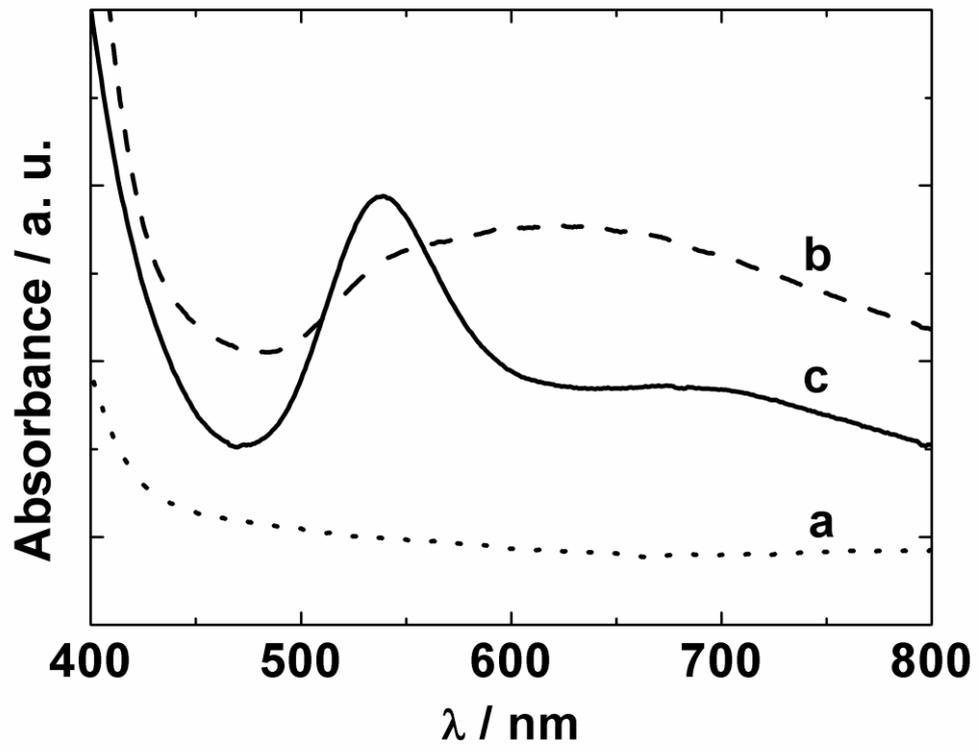

Fig. 4

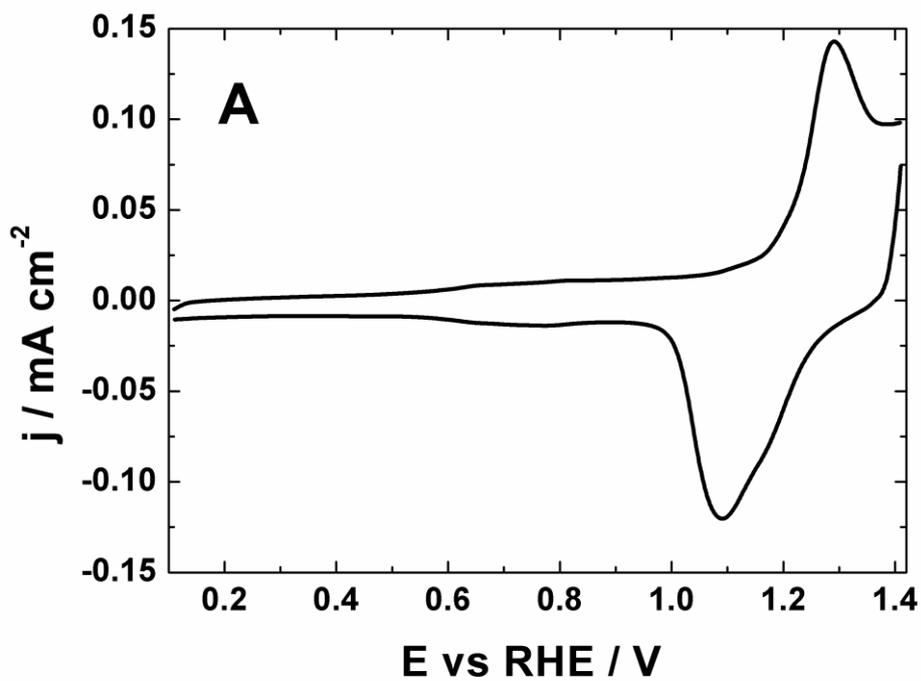

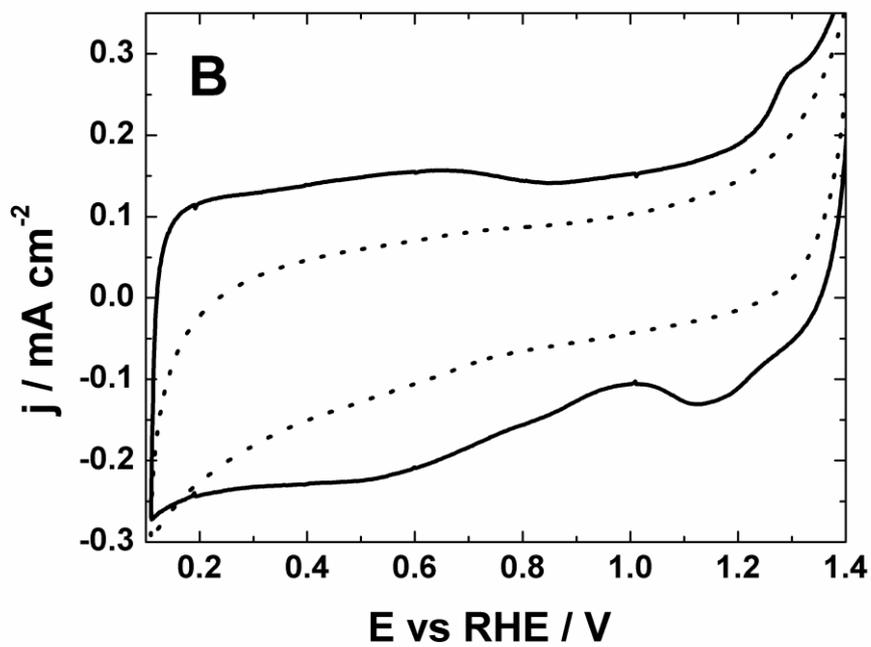

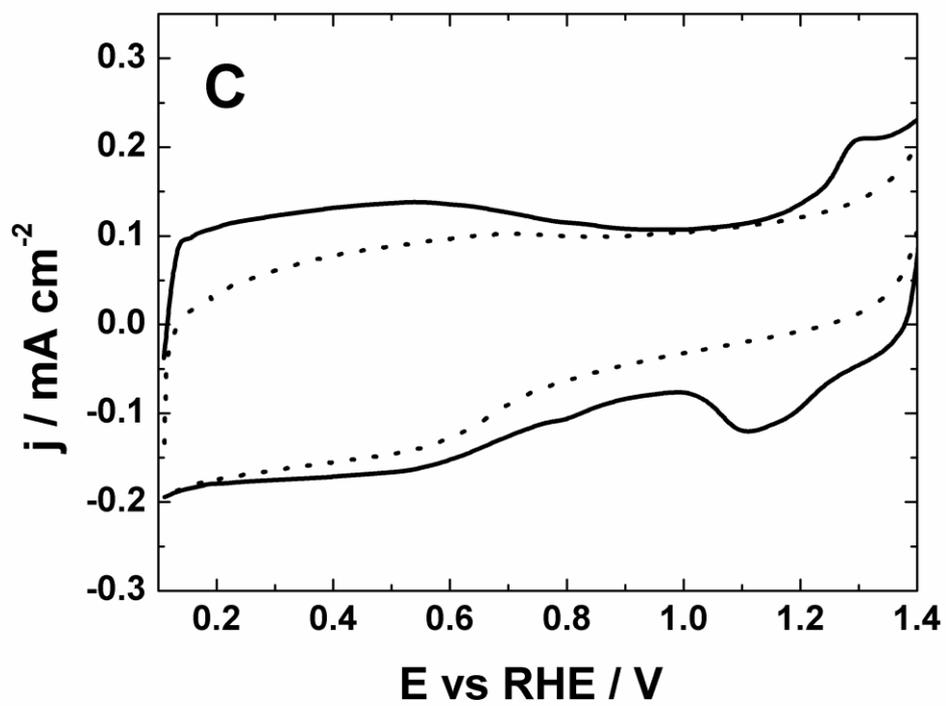

Fig. 5

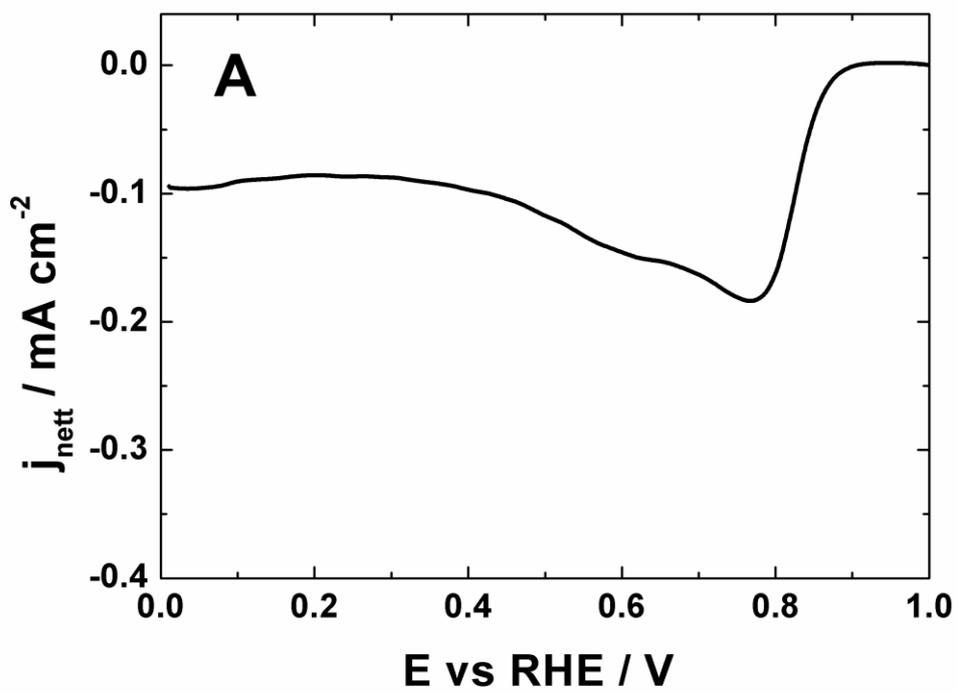

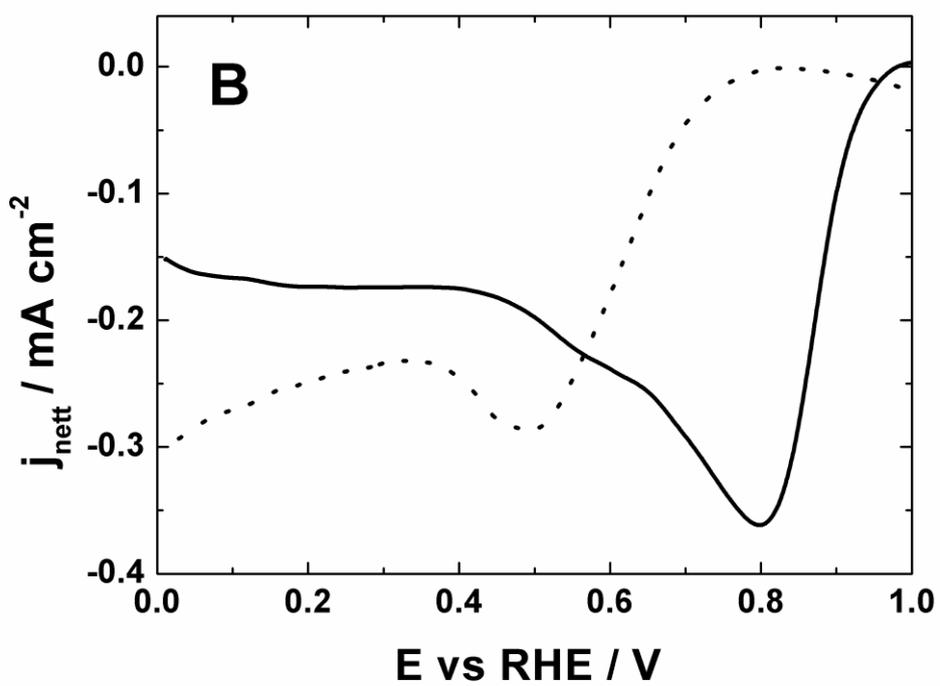

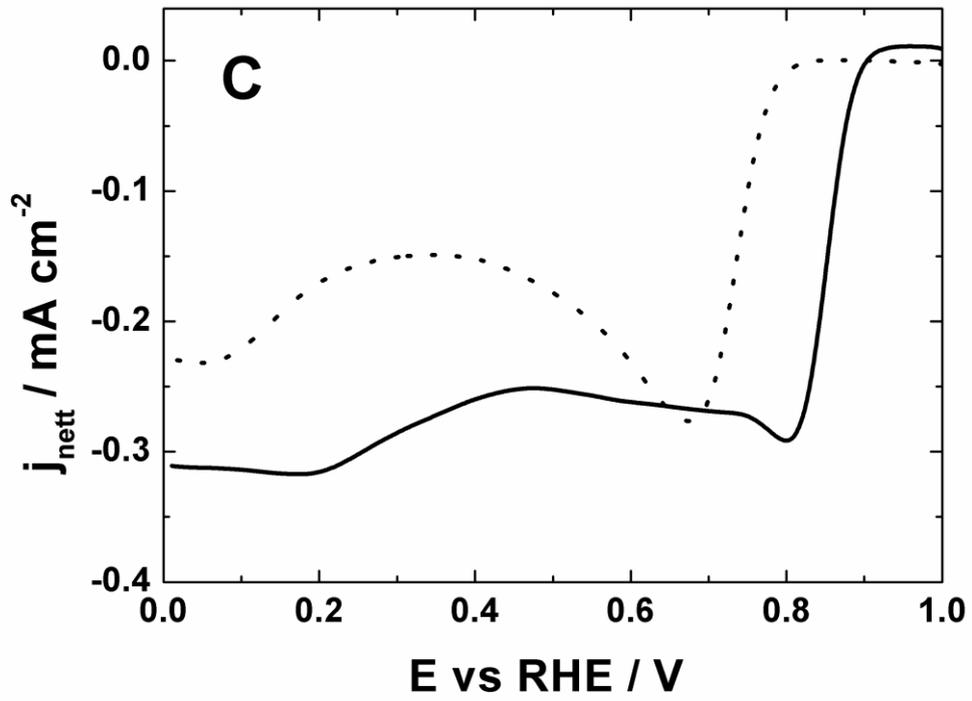

Fig. 6

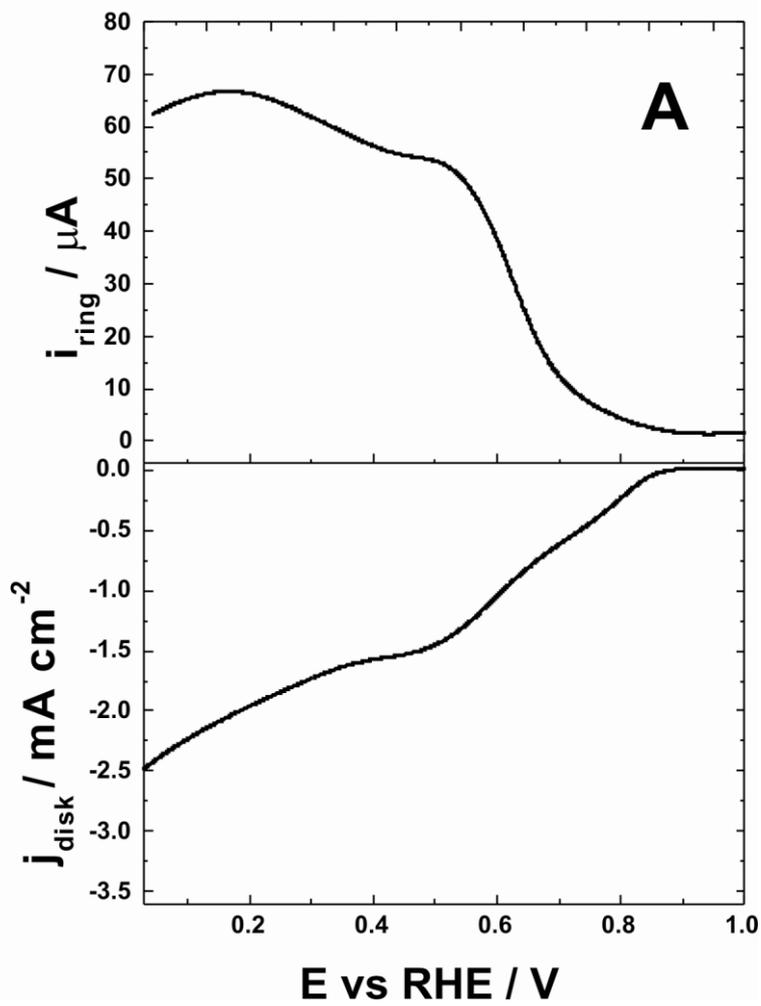

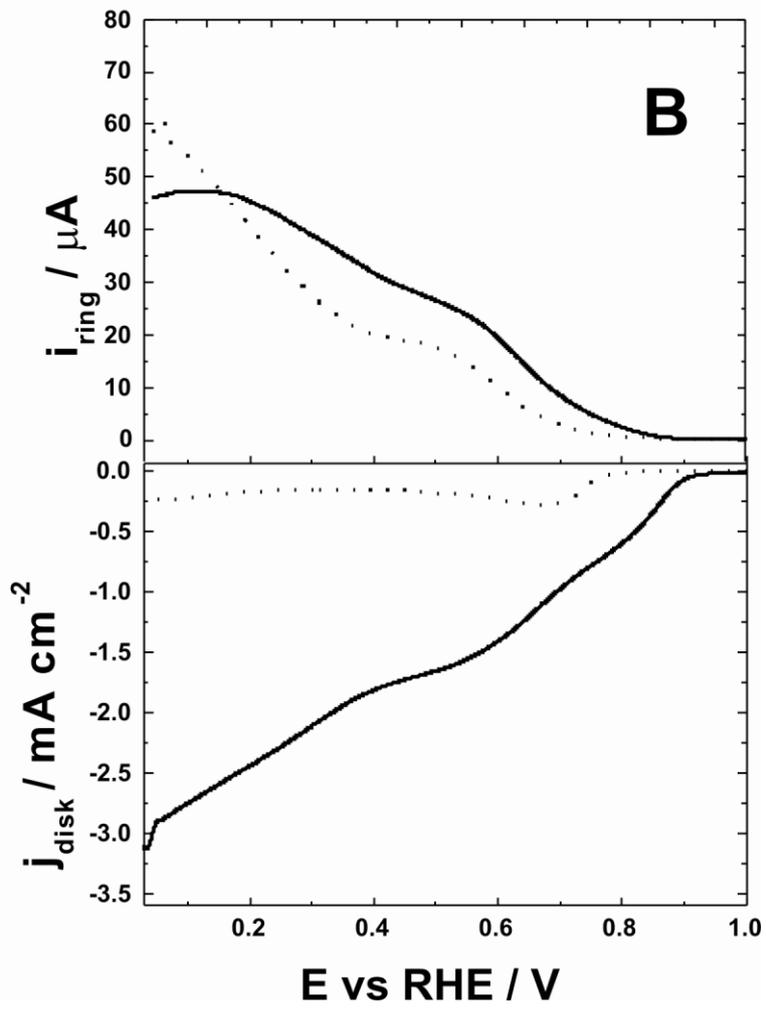

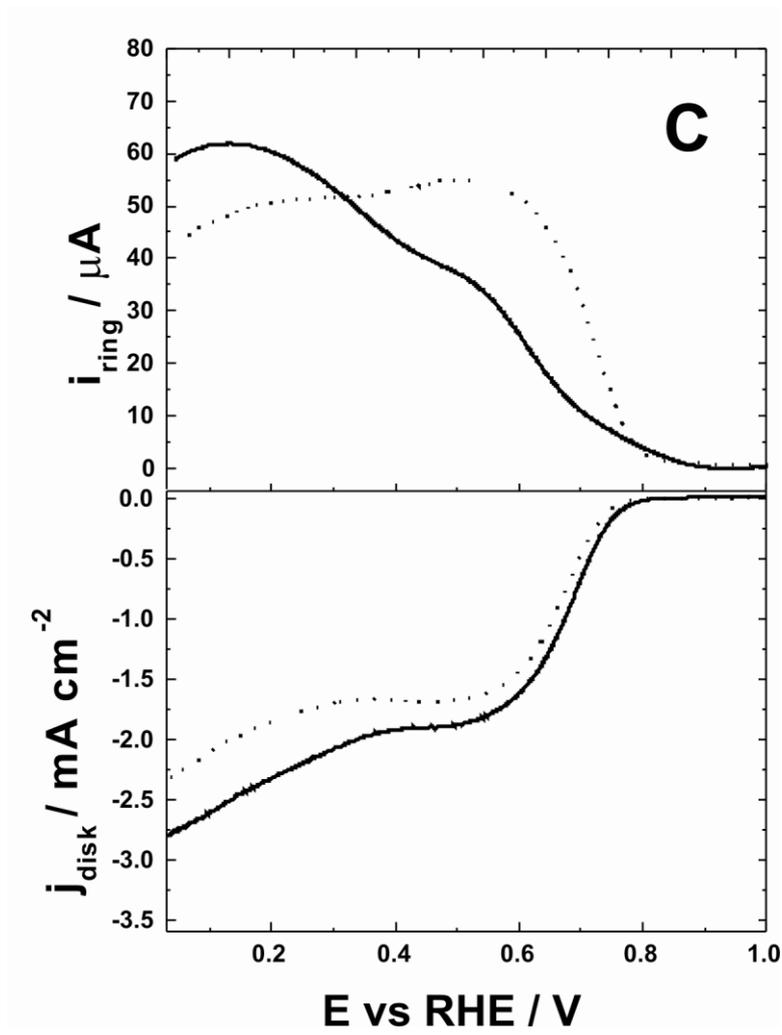

Fig. 7

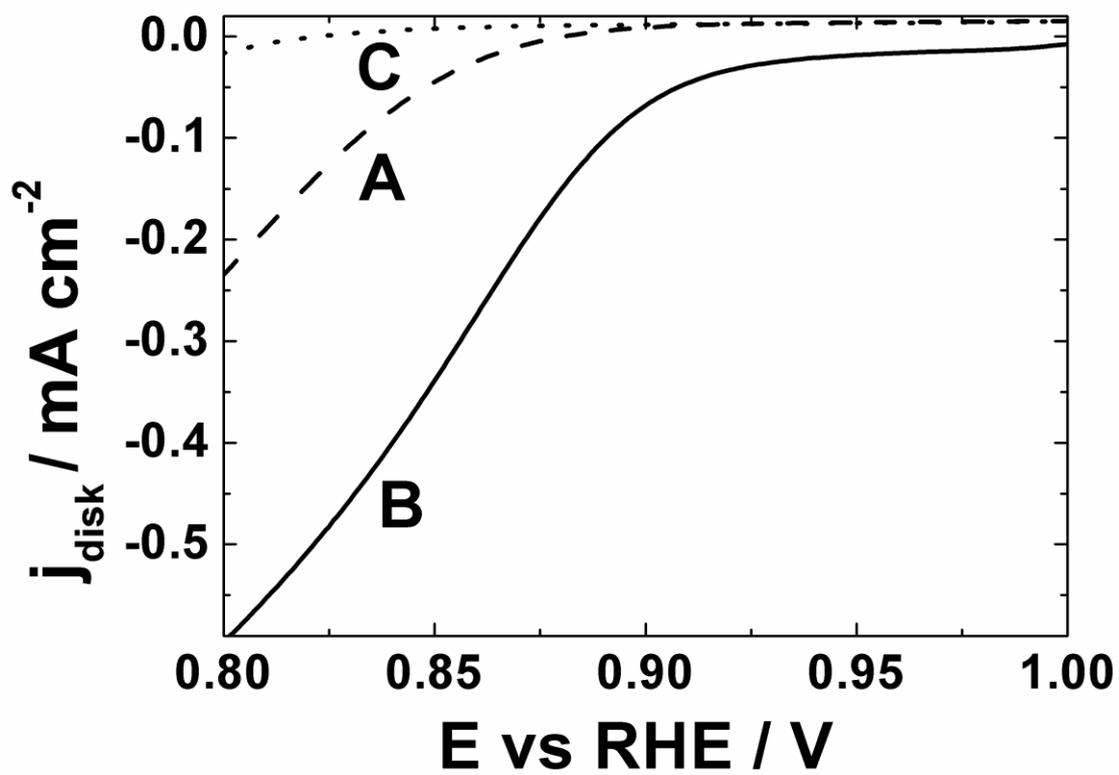

Fig. 8

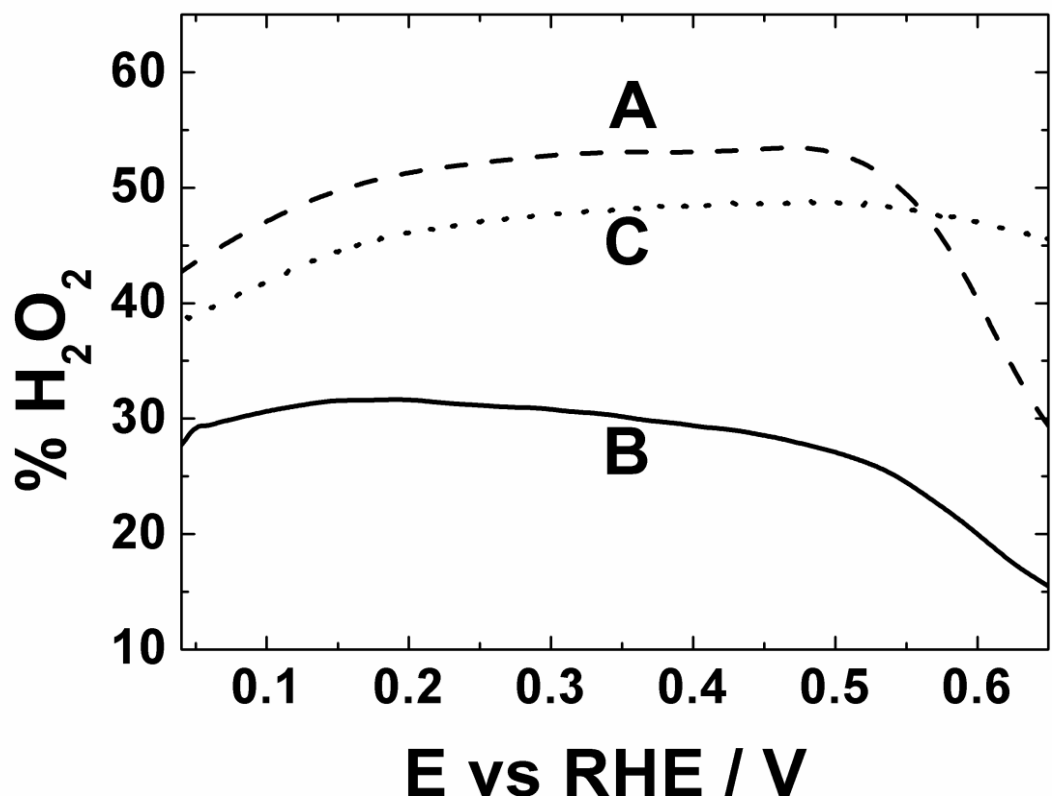

Fig. 9

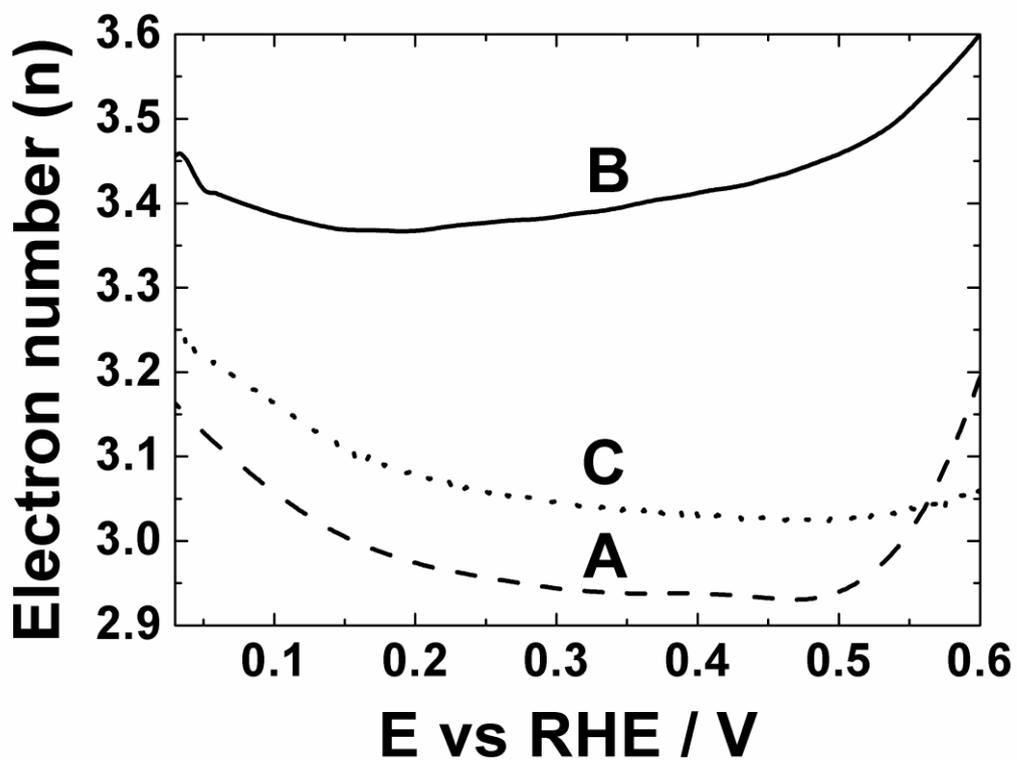

Fig. 10